\documentclass[letter]{aa} 
\usepackage{comment}
\usepackage{mathtools}
\usepackage{hyperref}
\hypersetup{
    colorlinks=true,
    linkcolor=blue,
    citecolor=blue,
    urlcolor=green,
    }
\usepackage{natbib}
\usepackage{graphicx}
\usepackage{txfonts}
\usepackage{ulem}
\usepackage{threeparttable}
\usepackage{mathrsfs}
\usepackage{amsmath}
\usepackage{placeins}
\usepackage{afterpage}
\usepackage{caption}
\begin{document}

\title{The spectrum of the persistent radio source associated with FRB~20190417A}

   \author{L. Bruno
          \inst{1},
          D. Pelliciari
          \inst{1},
          G. Bernardi
          \inst{1,2,3},
          M. Pilia
          \inst{4},
        L. Beduzzi
        \inst{1,5},
        P. Esposito
        \inst{6}
          }

   \institute{
       Istituto Nazionale di Astrofisica - Istituto di Radioastronomia, via Gobetti 101, 40129 Bologna, Italy  
      \and
    South African Radio Astronomy Observatory, Cape Town 7700, South Africa 
    \and
    Centre for Radio Astronomy Techniques and Technologies (RATT), Department of Physics and Electronics, Rhodes University, Makhanda 6140, South Africa
        \and
        Istituto Nazionale di Astrofisica - Osservatorio Astronomico di Cagliari, via della Scienza 5, I-09047, Selargius (CA), Italy
    \and
   Dipartimento di Fisica e Astronomia - Universit\`a di Bologna, via Gobetti 93/2, 40129 Bologna, Italy
          \and
    Scuola Universitaria Superiore IUSS Pavia, Palazzo del Broletto, piazza della Vittoria 15, I-27100 Pavia, Italy
    \\
    \email{l.bruno@ira.inaf.it}
 }
%   \date{Received September 15, 1996; accepted March 16, 1997}

% \abstract{}{}{}{}{} 
% 5 {} token are mandatory
 
  \abstract
  % context heading (optional)
   {Persistent radio sources (PRSs) are (sub-)parsec-scale compact non-thermal continuum sources associated with some repeating fast radio bursts (FRBs). Their nature is debated, but their properties provide insights into the FRB environment and progenitors.} 
  % aims heading (mandatory)
   {We measure the spectrum of the recently confirmed PRS associated with FRB 20190417A. Spectral features such as the self-absorption and cooling break can be used to constrain the age and size of PRSs and test theoretical models.}
  % methods heading (mandatory)
   {We present observations made with the 1.26 GHz upgraded Giant Metrewave Radio Telescope (uGMRT) and observations from the 6 GHz Karl Jansky Very Large Array (VLA). With complementary archival data and the LOw Frequency ARray Two Meter Sky Survey (LoTSS), we characterise the spectrum of the PRS between 144~MHz and 6~GHz. }
  % results heading (mandatory)
{The spectrum follows a power-law behaviour at gigahertz frequencies. The source is not detected at 144~MHz down to a $2\sigma=170 \; {\rm \mu Jy}$ sensitivity. We modelled the spectrum with a broken power law, obtaining a spectral index $\alpha = 0.20 \pm 0.05$ between 1-6 GHz. We placed a lower limit on the turn-over frequency of $> 370$~MHz ($95\%$ confidence). }
 % conclusions heading (optional)
 {The flat spectrum and low-frequency turn-over of the target are consistent with the spectral properties predicted for magneto-ionic nebulae, inflated behind the supernova ejecta by a flaring young magnetar. Considering the multi-zone magnetar wind nebula scenario, we estimate an age of $t< 250$~yr and a radius of $R< 0.4$~pc for the target, which would thus be slightly older than the PRSs associated with FRB~20121102A and FRB~20190520B.}


\titlerunning{The spectrum of the PRS associated with FRB~20190417A}
\authorrunning{Bruno et al.}
   \maketitle
%
%-------------------------------------------------------------------

\section{Introduction}

Fast radio bursts (FRBs; \citealt{lorimer07}) are bright (Jansky-level) flashes with a short duration ($\sim 1$  ms) mostly originating at extragalactic distances, whose nature is still debated (e.g. \citealt{Cordes&Chatterjee19,Petroff19,Bailes22,Zhang23FRB}, for reviews). Thousands of FRBs are known today, with a small fraction ($\sim 2\%$) of them showing repeating activity \citep{CHIME/FRBCollaboration19,CHIME26CAT2}. It is unclear whether the dichotomy between repeating and non-repeating FRBs is genuine or is the result of observational limits, implying that all FRBs can repeat on sufficiently long timescales (e.g. \citealt{Chime/FrbCollaboration23,Kirsten24}).

A handful of repeating FRBs (FRB 20121102A, FRB 20190520B, FRB 20240114A, and FRB 20190417A) is spatially coincident with persistent radio sources (PRSs), which are compact objects on (sub-)parsec scales emitting continuum non-thermal radiation
(\citealt{Chatterjee17,Marcote17,Niu22,Bhandari23,Ibik24,Bhusare25,bruni25,Moroianu26}). A candidate PRS is also associated with FRB 20201124A \citep{Bruni24}, but its size is poorly constrained ($<700$ pc). 

The engine powering PRSs and the mechanisms triggering FRBs are still unclear. A linear correlation was observed  between the rotation measure (RM) of the FRB and the spectral luminosity ($L_\nu$) of the associated PRS \citep{yang20,Bruni24}, which suggests that relativistic electrons powering synchrotron emission and thermal electrons causing Faraday rotation coexist within a strongly magnetised environment. In this framework, magnetar wind nebulae (MWNe; \citealt{Margalit&Metzger18}) and nebulae arising from hyper-accreting X-ray binaries \citep{Sridhar22} have been proposed as promising PRS engines. Alternatively, PRSs might be associated with core-dominated active galactic nuclei (AGN) from accreting intermediate-mass ($M\sim 10^2-10^5 \; M_{\odot}$) black holes (IMBHs; e.g. \citealt{Anna-Thomas23,Dong24}).

The spectral properties (spectral index, turn-over, and break frequency) can be used to constrain the origin of PRSs \citep[e.g.][]{Vohl23,Balasubramanian25,Bhattacharya25,Rahaman25}. In this Letter, we report on spectral measurements of the PRS associated with FRB 20190417A \citep{Ibik24,Moroianu26} by employing upgraded Giant Metrewave Radio Telescope (uGMRT) and Karl Jansky Very Large Array (VLA) data covering a frequency range of 1-8 GHz, and LOw Frequency ARray (LOFAR) data at 144 MHz.

Throughout this paper, we adopt a standard $\Lambda$CDM cosmology with $H_0=70\;\mathrm{km\; s^{-1}\; Mpc^{-1}}$, $\Omega_{\rm M}=0.3$, and $\Omega_{\rm \Lambda}=0.7$. We adopt the convention on the spectral index $\alpha$ as defined from the flux density as $S_\nu \propto \nu^{-\alpha}$. The paper is organised as follows. In Sect. \ref{sect: FRB 20190417A} we summarise the literature on the target. In Sect. \ref{sect: Data processing} we present the analysed radio data. In Sect. \ref{sect: Results} we derive the spectral properties of the PRS. In Sect. \ref{sect: Conclusions} we discuss our results.

\section{The PRS associated with FRB 20190417A}
\label{sect: FRB 20190417A}

The source FRB 20190417A (${\rm RA}=19^{\rm h} \, 39^{\rm m} \, 05.89^{\rm s}$, ${\rm Dec} = 59^{\rm o} \, 19' \, 36.83''$) is a repeating FRB with high dispersion measure ${\rm DM}= 1378$~pc~cm$^{-3}$ \citep{Fonseca20,Michilli23}, whose position is known at milliarcsec precision \citep{Moroianu26}. The source shows a high rotation measure ${\rm RM} \sim  4500 \; {\rm rad \; m^{-2}}$ \citep{McKinven23} with  
$\sim 20\%$ variations on timescales of a few months \citep{Moroianu26}. The FRB is associated with
a dwarf ($M\sim 7.6\times10^7 \; M_{\odot}$) low-metallicity ($Z\sim 0.2 \; Z_{\odot}$) star-forming galaxy (${\rm SFR}\sim 0.2 \; M_{\odot}\;{\rm yr}^{-1}$) at redshift $z=0.12817$ \citep{Ibik24,Moroianu26}.  

A candidate PRS with a flux density of $S_{1.52}=173\pm 9 \; {\rm \mu Jy}$ and compact down to a resolution of $1.3'' \times 1.1''$ was reported by \cite{Ibik24} at 1.52~GHz. \cite{Moroianu26} confirmed the association with the FRB and the compactness down to a resolution of $0.033'' \times 0.031''$ (implying a size $<23 \; {\rm pc}$), measuring $S_{1.4}=190\pm 40 \; {\rm \mu Jy}$.

The field of FRB 20190417A is included in a sample of 24 FRBs observed in the continuum with uGMRT in band 5 (1060-1460 MHz) to search for possible PRSs (Pelliciari et al., in prep.). The associated PRS was also followed up with the VLA in C band (4-8 GHz) in B configuration.

\section{Observations and data processing}
\label{sect: Data processing}
\begin{table*}
\fontsize{9}{9}\selectfont
\centering
\caption[]{Details of the radio data.}
\label{tab: dati}
\begin{tabular}{ccccccccc}
\hline
\noalign{\smallskip}
Instrument & Band & Frequency & Observation date & On-source time & Project code & P.I. & Flux, phase calibrators  \\
 & & (GHz) & & (min) & & &  \\
\noalign{\smallskip}
\hline
\noalign{\smallskip}
${\rm VLA_{[A]}}$ & L & 1-2 & 25-Feb.-2021 & 40  & 20B-280 & S. P. Tendulkar &  3C286, J1944+5448  \\
uGMRT &  5 & 1.06-1.46 & 05-Jul.-2024 & 120  & $46\_035$ & G. Bernardi & 3C48, J1944+5448 \\
${\rm VLA_{[B]}}$ & C & 4-8 & 03-Oct.-2025 & 8  & 25B-255 & D. Pelliciari &  3C48, J1927+6117  \\

\noalign{\smallskip}
\hline
\end{tabular}
\begin{tablenotes}
                \centering
        \item   {\small \textbf{Notes}. VLA data in L and C band are recorded in 16 and 32 spectral windows, each with 64 channels with a width of 1 and 2 MHz, respectively; uGMRT data are recorded in a single spectral window with 2048 channels and a width of 195 kHz.}  
        \end{tablenotes}   
\end{table*}

In this section, we present new observations of the PRS associated with FRB 20190417A taken with the uGMRT at 1.26~GHz and with the VLA at 6~GHz. We also reprocessed one of the six VLA observations at L band (1-2 GHz) presented by \cite{Ibik24}. The details of the observations are listed in Table~\ref{tab: dati}.

 We carried out a standard data reduction with the Common Astronomy Software Applications ({\tt CASA}; \citealt{mcmullincasapaper07}) v. 6.7 for all observations. After flagging data corrupted by RFI, we obtained delay, bandpass, and amplitude solutions from the primary calibrator and complex gain solutions from the secondary calibrator (Table \ref{tab: dati}). The flux density scale was set accordingly with \cite{perley&butler17}, with uncertainties $<5\%$. We additionally performed a round of phase self-calibration for the uGMRT data to improve the image quality. Imaging was performed with {\tt WSClean} \citep{offringa14,offringa17} v. 3.6 with a multi-frequency synthesis algorithm.

In addition, we retrieved three observations from the VLA Sky Survey \citep[VLASS;][]{lacyVLASS20} at S band ($2-4$~GHz) in B configuration, with a resolution of  $\theta\sim 2.5''$. Each epoch has a $\sigma = 120 \; {\rm \mu Jy \; beam^{-1}}$ rms noise. After convolving these images with a $\theta=3''\times3''$ beam, we averaged them and reached $\sigma = 75 \; {\rm \mu Jy \; beam^{-1}}$ in the stacked image. 

Finally, the field of FRB 20190417A was covered by the Third Data Release of the LoFAR Two Meter Sky Survey (LoTSS-DR3; \citealt{Shimwell26}) at 120-168 MHz. The 144 MHz image at $\theta= 6''$ has a noise of $\sigma =  85 \; {\rm \mu Jy \; beam^{-1}}$.

\section{Results}
\label{sect: Results}

\begin{figure*}
        \centering

\includegraphics[width=0.33\textwidth]{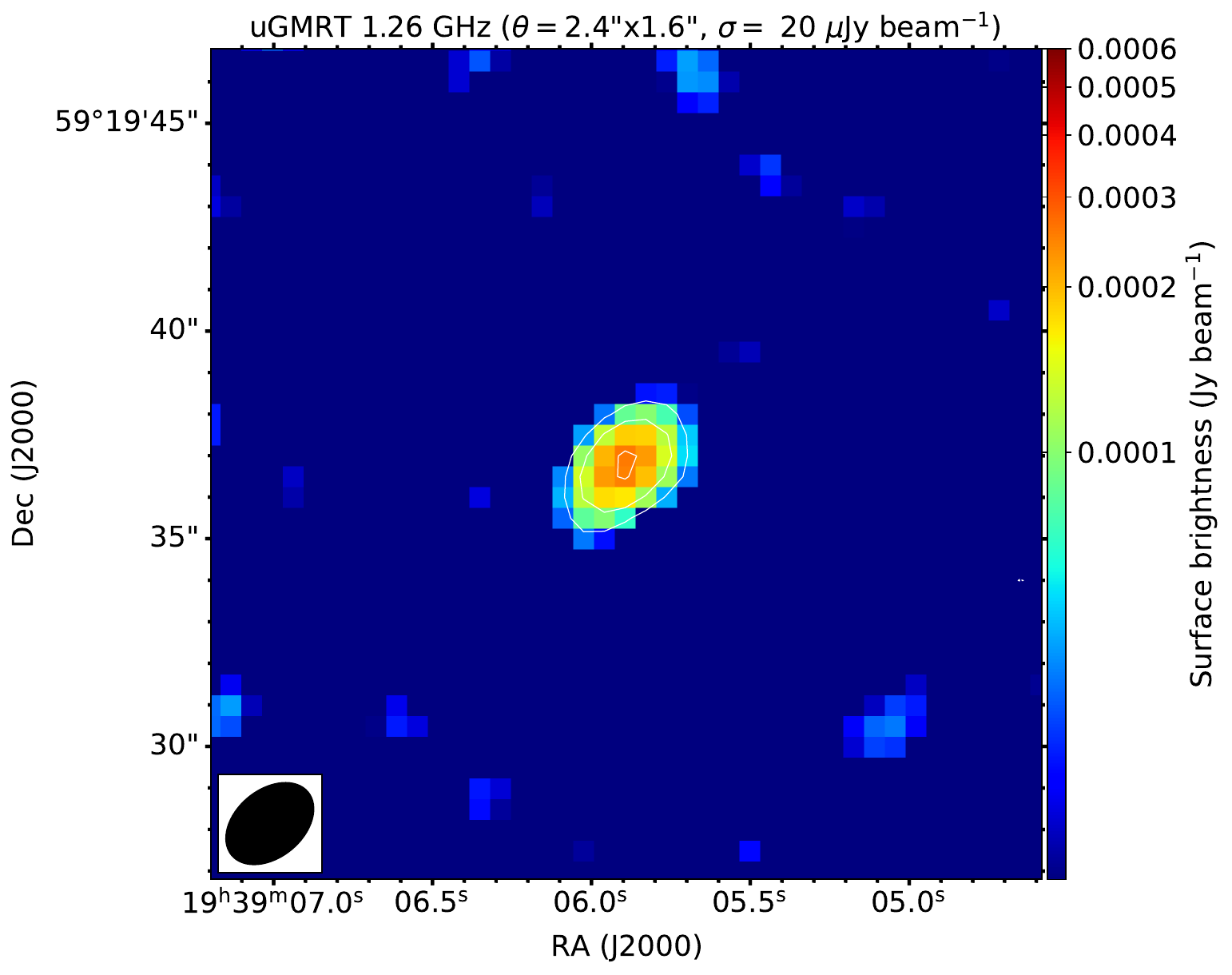}
\includegraphics[width=0.33\textwidth]{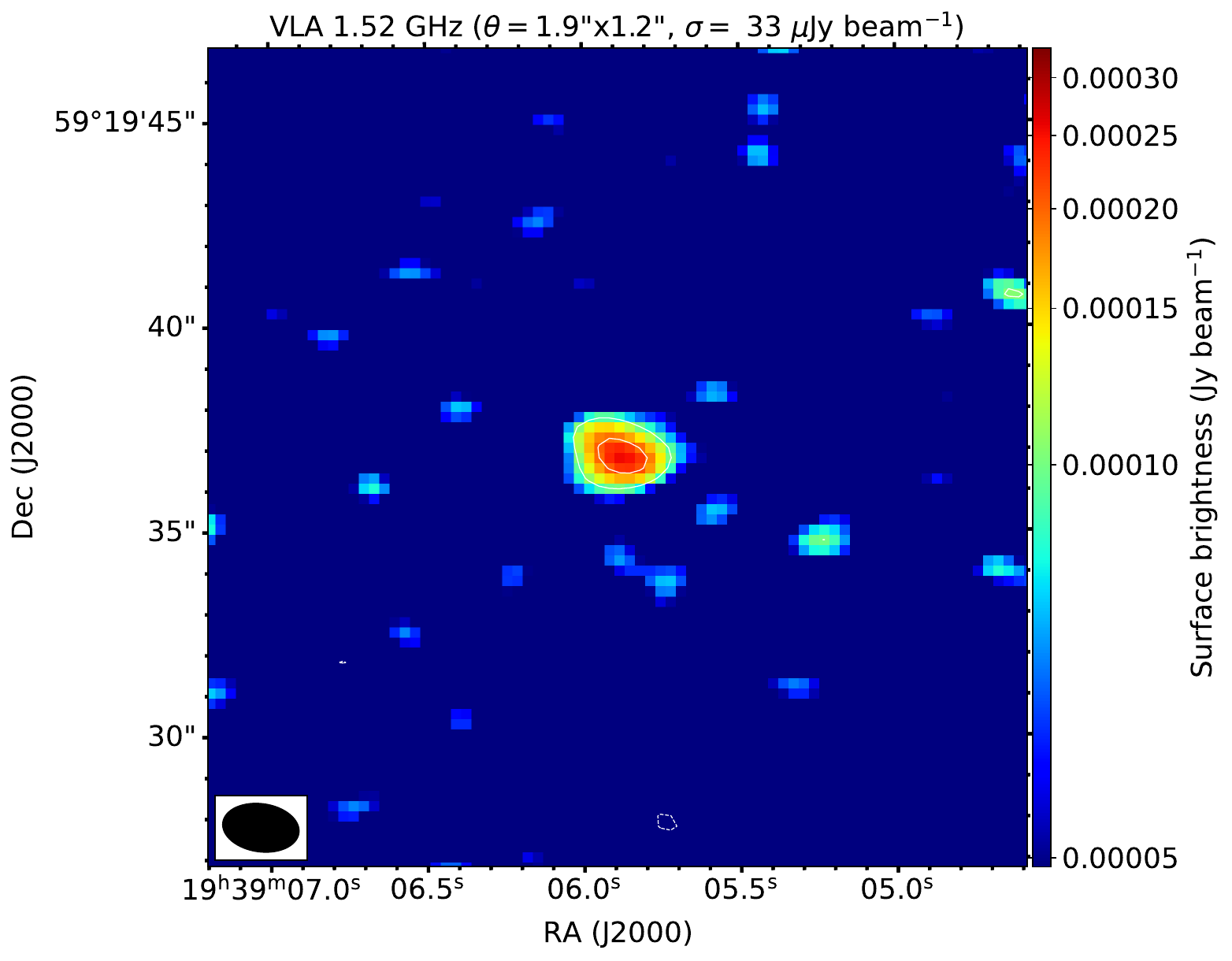}
\includegraphics[width=0.33\textwidth]{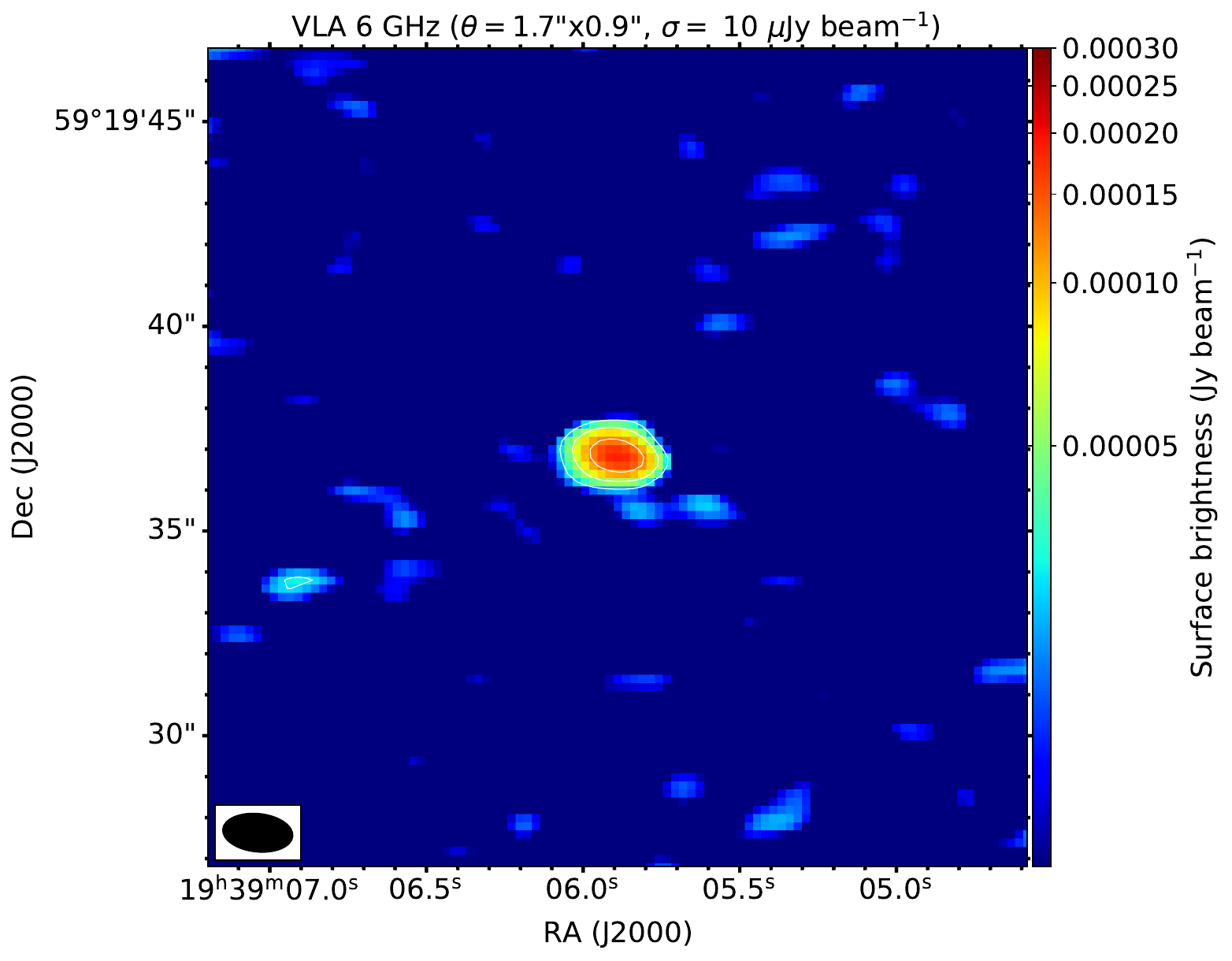}

        \caption{Images of the PRS. From left to right: uGMRT at 1.26 GHz, VLA at 1.52 GHz, and VLA at 6 GHz. The resolution and noise are reported at the top of each panel. The contour levels are $[\pm3, \;6, \;12]\times \sigma$. }
        \label{fig: maps}
\end{figure*}

\begin{table}
\fontsize{8.5}{8.5}\selectfont
\centering
        \caption[]{Properties of radio images shown in Figs. \ref{fig: maps}, \ref{fig: surveys} and corresponding flux densities of the PRS. }
        \label{tab: radio flux} 
% \resizebox{\textwidth}{!}{
 \begin{tabular}{cccccc}
        \hline
        \noalign{\smallskip}
        $\nu$ & Weight & $\theta$ & $PA$ & $\sigma$ & $S$    \\  
        (GHz) & & ($''\times''$) & (deg) & ($\mu \rm{Jy} \; beam^{-1}$) &  ($\mu$Jy)   \\  
    \hline
        \noalign{\smallskip}
        1.26 &  -0.5 & $2.4\times1.6$ & 50 & 20 & 268  \\
        1.52 &  0.5 & $1.9\times1.2$ & 98 & 33 & 248   \\
        3 &  - & $3.0\times3.0$ & 0 & 75 & 202   \\
        6 &  0.5 & $1.7\times0.9$ & 96 & 10 & 184   \\
\noalign{\smallskip}
        \hline
        \end{tabular}
 %}
\begin{tablenotes}
                \centering
        \item   {\small \textbf{Notes}. Cols. 1-6: Central frequency, robust parameter for Briggs \citep{briggs95} weighting, beam position angle (W to N), rms noise of the images, and measured peak flux densities of the PRS.}  
        \end{tablenotes}        
\end{table}

We show the uGMRT and VLA images in Fig. \ref{fig: maps} and summarise their properties in Table \ref{tab: radio flux}. The PRS is detected in all these images with a high signal-to-noise ratio (${\rm S/N}\sim 8-18$). A low-significance ($2\sigma$) peak is also detected in the VLASS image, but is not detected in the LoTSS image (Fig. \ref{fig: surveys}).

\subsection{Flux density measurements}
\label{sect: Flux density measurements}

We performed a Gaussian fit to the surface brightness of the PRS to obtain flux density measurements. The fitted peak values are reported in Table~\ref{tab: radio flux} (with uncertainties $\Delta S=\sigma$).

Our 1.52 GHz VLA flux density is higher by a factor of 1.3 than the 1.4 GHz value reported by \cite{Moroianu26} and derived from European Very Long Baseline Interferometry Network (EVN) observations. Although the two measurements are consistent within the uncertainties, this difference may reflect a contribution from extended emission, which remains unresolved in our arcsecond-resolution images, while it is not recovered by EVN due to its lower sensitivity at short spacings. On the other hand, we note that our $S_{1.52}$ measurement is inconsistent ($95\%$ confidence) with 
\cite{Ibik24}. The target is located $\sim 10'$ away from the pointing centre, where the primary beam attenuation is $\sim 65\%$ \citep{Perley2016JanskyVL}. When this correction is not accounted for, the two flux density measurements agree well, which might  suggest that \cite{Ibik24} did not properly account for it. As a further check, we measured  $S_{1.52}$ for ten compact sources in the field and compared it with $S_{1.4}$ from the National Radio Astronomy Observatory VLA Sky Survey (NVSS; \citealt{condon98NVSS}). The agreement was good, with a  mean $S_{1.4}$ to $S_{1.52}$ ratio of 1.0 (and a standard deviation of 0.3), which rules out any impacting flux scale offset in our image. In the following, we exclude $S_{1.52}$ by \cite{Ibik24} from our analysis.

\subsection{Radio spectrum and spectral luminosity}
\label{sect: Radio spectrum}

\begin{figure}
    \centering
    \includegraphics[width=0.49\textwidth]{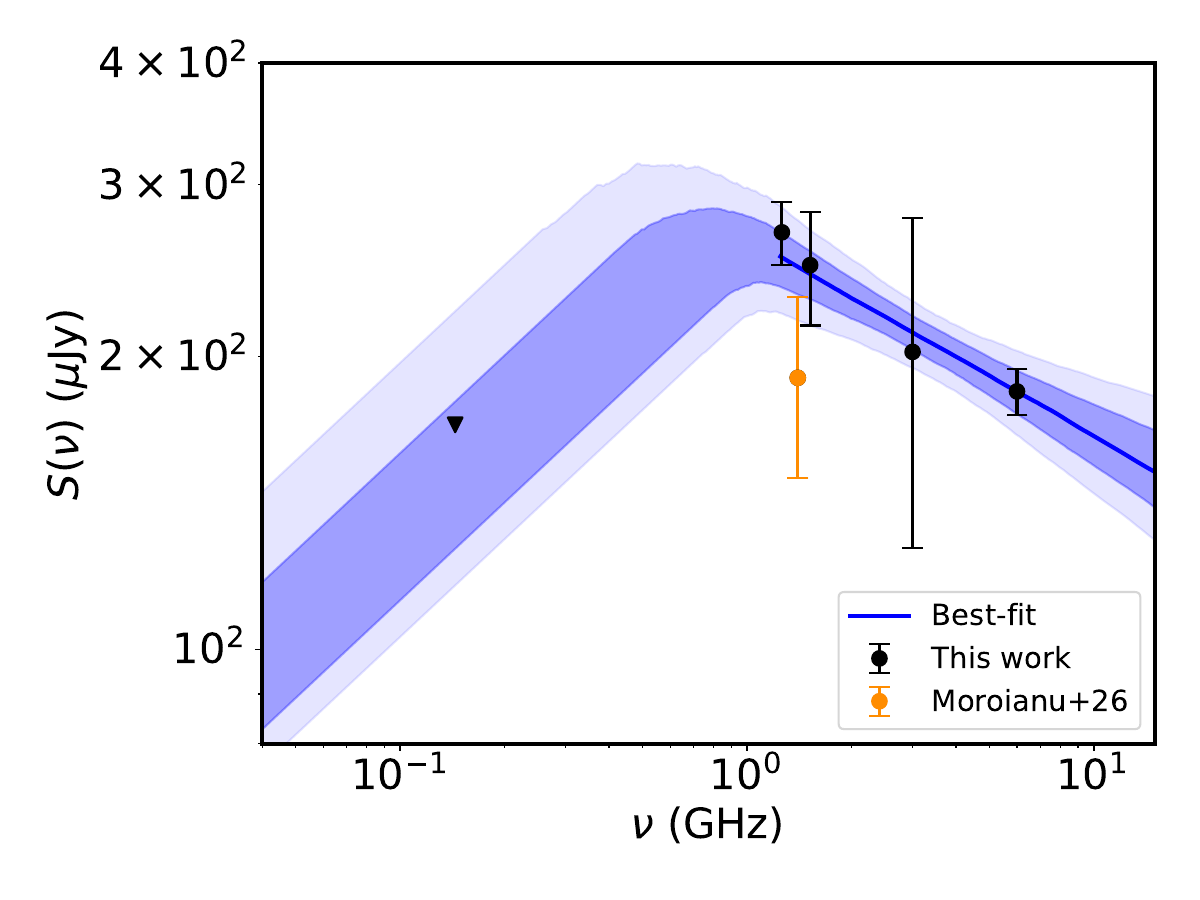}
    \caption{Radio spectrum of the PRS. The flux densities (dots) and upper limit (triangle) were fitted with a broken power law (Eq. \ref{eq: model spectrum}) through the MCMC method. The shaded regions indicate the $68\%$ (dark) and $95\%$ (light) confidence levels. The posterior probability distributions of the fitted parameters $(\nu_{\rm to},\; S_3, \; \alpha)$ are shown in Fig. \ref{fig: corner}.}
    \label{fig: S5_spectrum_break}
\end{figure}

The PRS flux densities (Table \ref{tab: radio flux} and $S_{1.4}$ by \citealt{Moroianu26}) and the $2\sigma$ upper limit provided by LoTSS ($S_{144}<170 \; \mu {\rm Jy} $) are shown in Fig. \ref{fig: S5_spectrum_break}. A single power law fits observations in the 1--6 GHz range well, but the extrapolation of the spectrum at 144 MHz exceeds the upper limit by a factor of $\sim 2$ . This suggests a turn-over ($\nu_{\rm to}$) at low frequency. 

To model the observed spectrum, we considered a broken power law describing two optically thin regimes dominated by low-energy ($S\propto \nu^{1/3}$) and high-energy ($S\propto \nu^{-\alpha}$) relativistic electrons (see Appendix \ref{sect: Spectral modelling} for details), in the form
\begin{equation}
S(\nu) =
\begin{cases}
S_3 \left(\dfrac{\nu}{\nu_{\rm to}}\right)^{1/3} \left(\dfrac{\nu_{\rm to}}{3\,\mathrm{GHz}}\right)^{-\alpha},
& \nu < \nu_{\rm to} \\[1.2em]
S_3 \left(\dfrac{\nu}{3\,\mathrm{GHz}}\right)^{-\alpha},
& \nu \ge \nu_{\rm to},
\end{cases}
\label{eq: model spectrum}
\end{equation}
where the normalisation was set at 3 GHz as the (approximate) average frequency between 1.2 and 6~GHz. Following a Markov chain Monte Carlo (MCMC) approach, we fitted the spectrum in Eq. \ref{eq: model spectrum}, leaving $S_3$, $\nu_{\rm to}$, and $\alpha$ free to vary within uniform prior ranges (Appendix \ref{sect: Spectral modelling}). 

The spectral index at gigahertz frequencies is well constrained, with a best-fit value of $\alpha=0.20\pm0.05$. While this result differs from the steep spectrum ($\alpha\sim 1.2$) reported by \cite{Ibik24}, their measurement was derived within a narrower frequency range (1-2 GHz). We found a best-fit flux density $S_3=211\pm 8 \; {\rm \mu Jy}$ at 3~GHz, whereas we set an upper limit $\nu_{\rm to} \geq 370$ MHz ($95\%$ confidence level) to the turn-over frequency. We stress that the high-frequency spectral index is robust against the choice of the upper-limit significance, while $\nu_{\rm to}$ is more sensitive to this assumption. Specifically, a more conservative $3\sigma$ upper limit at 144 MHz would result in slightly lower turn-over frequency values ($\nu_{\rm to} \geq 300$ MHz), but this difference relative to the $2\sigma$ case remains marginal. 

By assuming the best-fit spectral index, we derived the $k$-corrected spectral luminosity as $L_{\rm \nu}= 4 \pi D_{\rm L}^{2}S_{\rm \nu}(1+z)^{{\alpha-1}}$, where $D_{\rm L}=601$ Mpc is the luminosity distance. This yields a 1.5 GHz spectral luminosity of $L_{\rm 1.5}= (9.5\pm 0.8)\times 10^{28} \; {\rm erg \; s^{-1} \; Hz^{-1}}$.

\section{Discussion and conclusions}
\label{sect: Conclusions}

We presented the spectrum of the PRS associated with FRB 20190417A. At 1.26--6 GHz, the spectrum follows a power law with a flat spectral index $\alpha=0.20\pm 0.05$. The PRS is not detected at 144~MHz ($S_{\rm 144}<170 \; {\rm \mu Jy}$), showing a turn-over at low frequencies. We estimated the turn-over frequency to be in the range of 0.37--1.26 GHz. The source spectral luminosity at 1.5 GHz is $L_{\rm 1.5}= (9.5\pm 0.8)\times 10^{28} \; {\rm erg \; s^{-1} \; Hz^{-1}}$.

The PRSs associated with FRB 20121102A (PRS1), FRB 20190520B (PRS2), FRB 20240114A (PRS3), and FRB 20190417A (PRS4) reside in dwarf star-forming galaxies with low metallicity \citep{Bassa17,Niu22,bruni25,Moroianu26}. These tend to exhibit flat spectra ($\alpha \sim 0.1-0.4$) at gigahertz frequencies and luminosities in the range $L_{\rm \nu}\sim 10^{28}-10^{29}\,{\rm erg\,s^{-1}\,Hz^{-1}}$. Notably, PRS2 shows a turn-over below $\sim 1$ GHz
\citep{Balasubramanian25}, similarly to PRS4. A turn-over might also exist for PRS3, but the prominent flux density variability on short timescales \citep{Zhang25} prevents a firm conclusion. The spectrum of PRS1 steepens at $\sim 10-20$ GHz ($\alpha\sim 1$), whereas current low-frequency data down to 400 MHz do not provide solid evidence for a turn-over \citep{Resmi21,Bhardwaj25}. 

A low-frequency turn-over, caused by synchrotron self-absorption, is predicted by the MWN model \citep{Margalit&Metzger18}. The MWN multi-zone model described in \cite{Rahaman25} assumes an expanding magnetised nebula surrounded by a supernova remnant and embedding a young ($\sim 10^2$ yr) magnetar with a long period ($\sim 10$ ms). This model predicts a self-absorption frequency at $\sim 200-500$ MHz for PRS1 and PRS2. The flat spectrum of PRS4, similar to that of pulsar wind nebulae \citep[e.g.][]{Gaensler&Slane06,Kothes17}, and its low-frequency turn-over are consistent with the MWN model.

In conclusion, a MWN scenario is viable for PRS4. We used scaling relations from \citealt{Rahaman25} (see Appendix \ref{sect: Spectral modelling}) to provide limits on the radius and age of PRS4, finding $R<0.4$ pc (consistent with the current limit from \citealt{Moroianu26}) and $t< 250$ yr. In this scenario, PRS4 would thus be slightly older and more extended than PRS1 and PRS2. We note that \cite{Moroianu26} argued that the hosts of known PRSs are thought to favour the formation of short-period ($\sim 3$ ms) magnetars, thus challenging the assumption of long-period magnetars in the model proposed by \cite{Rahaman25}. This might either indicate limitations of the multi-zone MWN model by \cite{Rahaman25} or the need of considering alternative scenarios, such as a hyper-accreting X-ray binary system \citep{Sridhar22} or low-power and core-dominated AGN from IMBHs \citep{Eftekhari20,Anna-Thomas23,Dong24,Bhardwaj25}.

\begin{acknowledgements}
The authors thank the referee for their comments and suggestions. LB thanks A. Ibik for kindly sharing her images. The research activities described in this paper were carried out with contribution of the NextGenerationEU funds within the National Recovery and Resilience Plan (PNRR), Mission 4 - Education and Research, Component 2 - From Research to Business (M4C2), Investment Line 3.1 - Strengthening and creation of Research Infrastructures, Project IR0000026 – Next Generation Croce del Nord. The National Radio Astronomy Observatory and Green Bank Observatory are facilities of the U.S. National Science Foundation operated under cooperative agreement by Associated Universities, Inc. We thank the staff of the GMRT that made these observations possible. GMRT is run by the National Centre for Radio Astrophysics of the Tata Institute of Fundamental Research. LOFAR \citep{vanhaarlem13} is the Low Frequency Array designed and constructed by ASTRON. It has observing, data processing, and data storage facilities in several countries, which are owned by various parties (each with their own funding sources), and that are collectively operated by the ILT foundation under a joint scientific policy. The ILT resources have benefited from the following recent major funding sources: CNRS-INSU, Observatoire de Paris and Universit\'e d’Orl\'eans, France; BMBF, MIWF- NRW, MPG, Germany; Science Foundation Ireland (SFI), Department of Business, Enterprise and Innovation (DBEI), Ireland; NWO, The Netherlands; The Science and Technology Facilities Council, UK; Ministry of Science and Higher Education, Poland; The Istituto Nazionale di Astrofisica (INAF), Italy. This research made use of the Dutch national e-infrastructure with support of the SURF Cooperative (e-infra 180169) and the LOFAR e-infra group. The J\"ulich LOFAR Long Term Archive and the German LOFAR network are both coordinated and operated by the J\"ulich Supercomputing Centre (JSC), and computing resources on the supercomputer JUWELS at JSC were provided by the Gauss Centre for Supercomputing e.V. (grant CHTB00) through the John von Neumann Institute for Computing (NIC). This research made use of the University of Hertfordshire high-performance computing facility and the LOFAR-UK computing facility located at the University of Hertfordshire and supported by STFC [ST/P000096/1], and of the Italian LOFAR-IT computing infrastructure supported and operated by INAF, including the resources within the PLEIADI special `LOFAR' project by USC-C of INAF, and by the Physics Department of Turin University (under the agreement with Consorzio Interuniversitario per la Fisica Spaziale) at the C3S Supercomputing Centre, Italy. This research made use of SAOImageDS9, developed by Smithsonian Astrophysical Observatory \citep{ds9}, APLpy, an open-source plotting package for Python \citep{robitaille&bressert12APLPY}, Astropy, a community-developed core Python package for Astronomy \citep{astropycollaboration13,astropycollaboration18}, Matplotlib \citep{hunter07MATPLOTLIB}, Numpy \citep{harris20NUMPY}, SciPy \citep{scipy}.

\end{acknowledgements}

\bibliographystyle{aa}
\bibliography{bibliografia}

\begin{appendix}

\FloatBarrier

\section{LoTSS and VLASS images}
\label{sect: VLASS and LoTSS images}

\begin{figure}[!htb]
        \centering
        \includegraphics[width=0.33\textwidth]{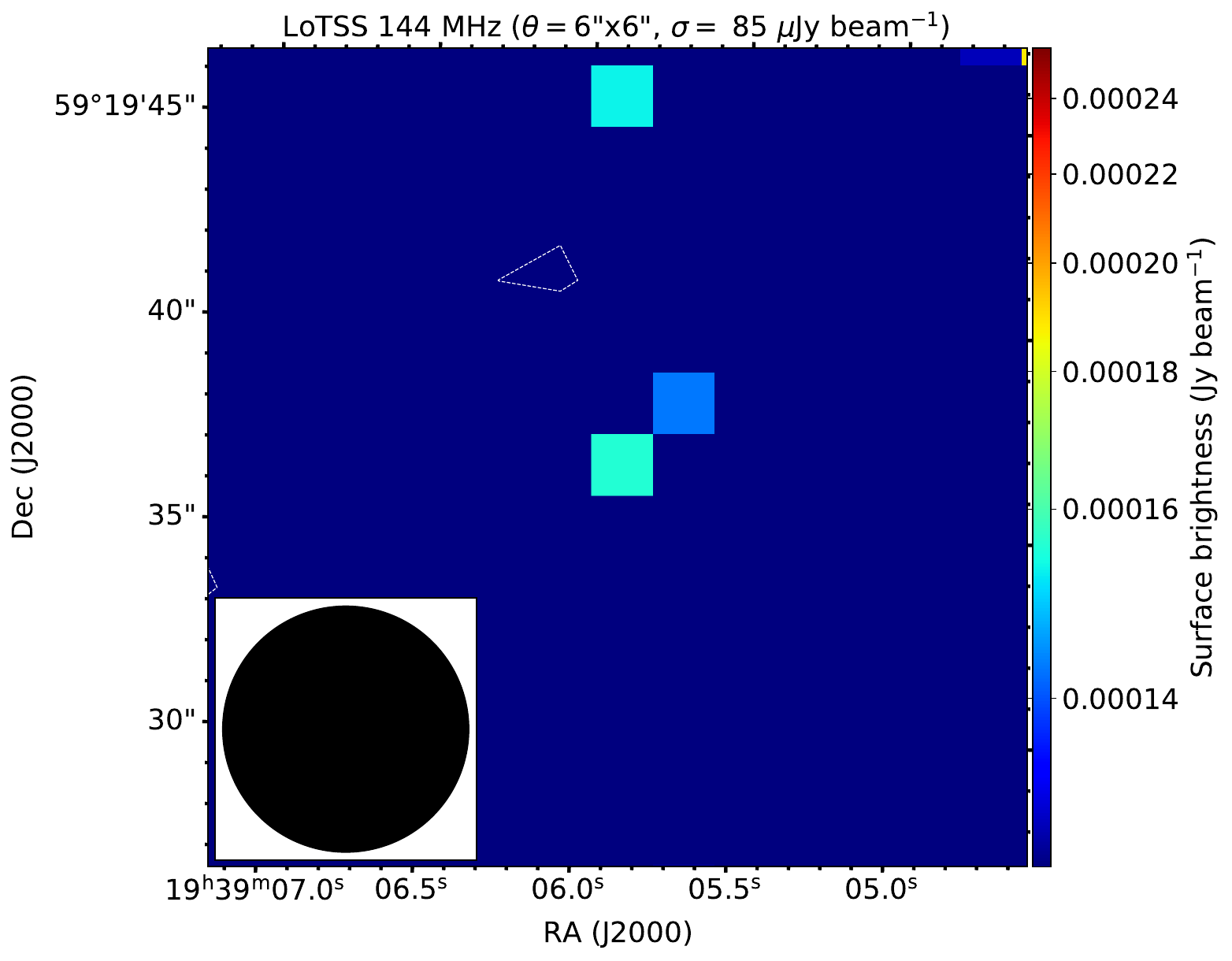}
\includegraphics[width=0.33\textwidth]{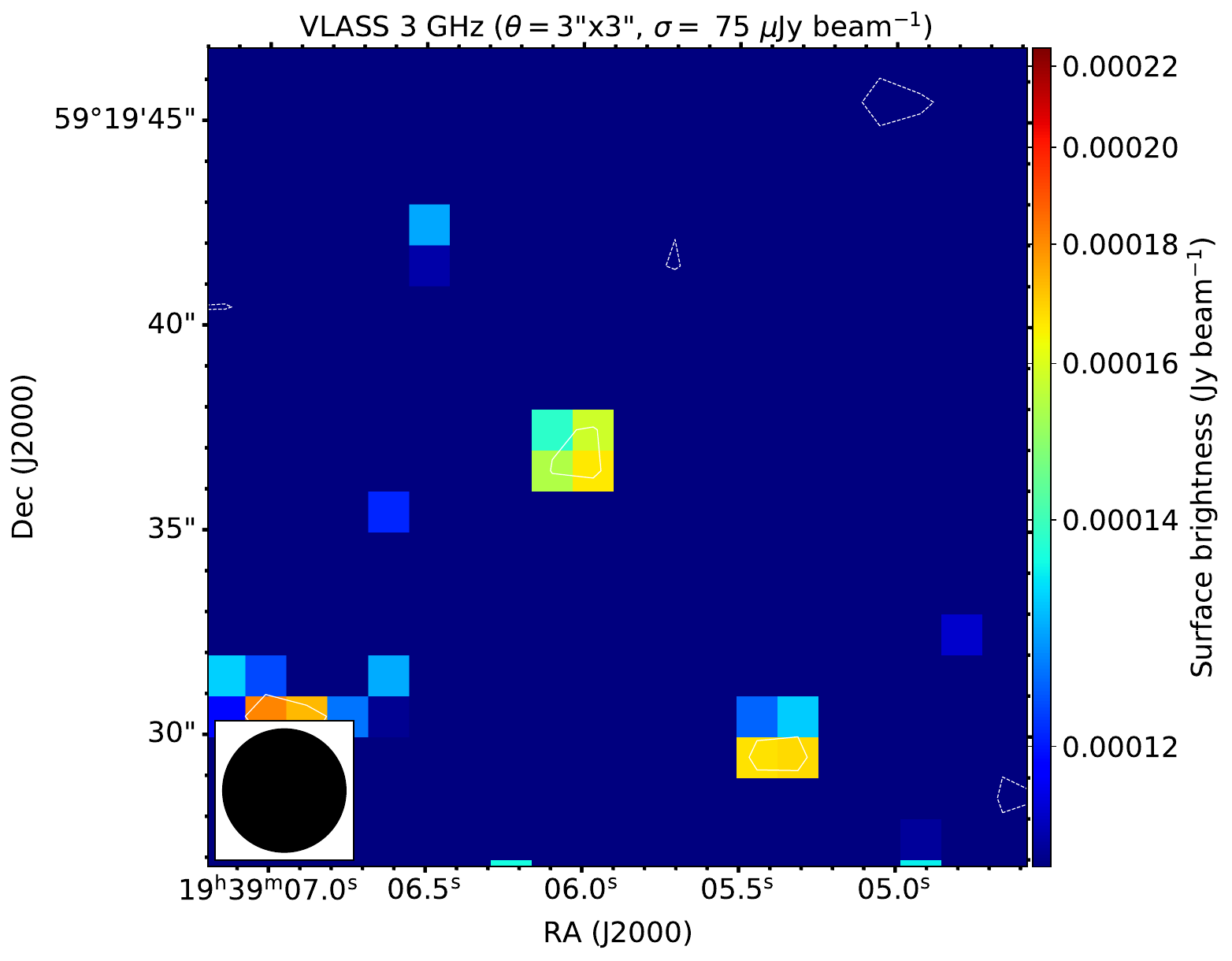}

        \caption{Images from LoTSS at 144 MHz and VLASS at 3 GHz covering the field of the PRS. The resolution and noise are reported on top of each panel. The contour levels are $[\pm2]\times \sigma$.}
        \label{fig: surveys}
\end{figure}

In Fig. \ref{fig: surveys} we show the LoTSS-DR3 and stacked VLASS images used in this work. The PRS is barely detected at 3 GHz and is undetected at 144 MHz. 

\FloatBarrier

\section{Spectral modelling}
\label{sect: Spectral modelling}

The spectral model of the PRSs associated with FRB 20121102A and FRB 20190520B considered by \cite{Rahaman25} to describe an expanding nebula surrounding the magnetar is based on spectrum '1' presented in Fig. 1 in \cite{Granot&Sari02}. This spectrum includes four regimes determined by the self-absorption ($\nu_{\rm sa}$), peak ($\nu_{\rm m}$), and cooling break ($\nu_{\rm c}$) frequencies. These are i) a self-absorbed, optically thick, regime ($\nu <\nu_{\rm sa}$) where $S\propto \nu^2$, ii) an optically thin regime ($\nu_{\rm sa}<\nu <\nu_{\rm m}$) dominated by low-energy relativistic electrons where $S\propto \nu^{\frac{1}{3}}$, iii) an optically thin regime ($\nu_{\rm m}<\nu <\nu_{\rm c}$) dominated by high-energy relativistic electrons where $S\propto \nu^{-\alpha}$, and iv) a cooling-dominated regime ($\nu >\nu_{\rm c}$) where $S\propto \nu^{-\alpha - 0.5}$. These frequencies are function of time, and they progressively shift towards lower values following the expansion of the nebula. 

We aim to test the model proposed by \cite{Rahaman25} for our target. As we have a limited number of data points, we considered a simplified model (Eq. \ref{eq: model spectrum}) with only two regimes defined by a turn-over frequency. Although simplistic, this model is motived by available measurements, as we have no evidence of the cooling break at GHz frequencies (implying $\nu_{\rm c} > 6 \; {\rm GHz}$), while at lower frequencies ($<1 \; {\rm GHz}$) we have only an upper limit. We stress that this upper limit (rather than a flux density measurement) does not allow us to distinguish between the self-absorbed and optically thin regimes, but in our model we assumed the latter because the coincidence of $\nu_{\rm sa}$ and $\nu_{\rm m}$ is unlikely (this condition verifies only for a short duration transition). Qualitatively, adopting a shallower slope ($S\propto \nu^{\frac{1}{3}}$) provides a more restrictive lower limit on $\nu_{\rm to}$, whereas $S\propto \nu^2$ would enable the turn-over to occur at lower frequencies.

\begin{figure}
    \centering
    \includegraphics[width=0.45\textwidth]{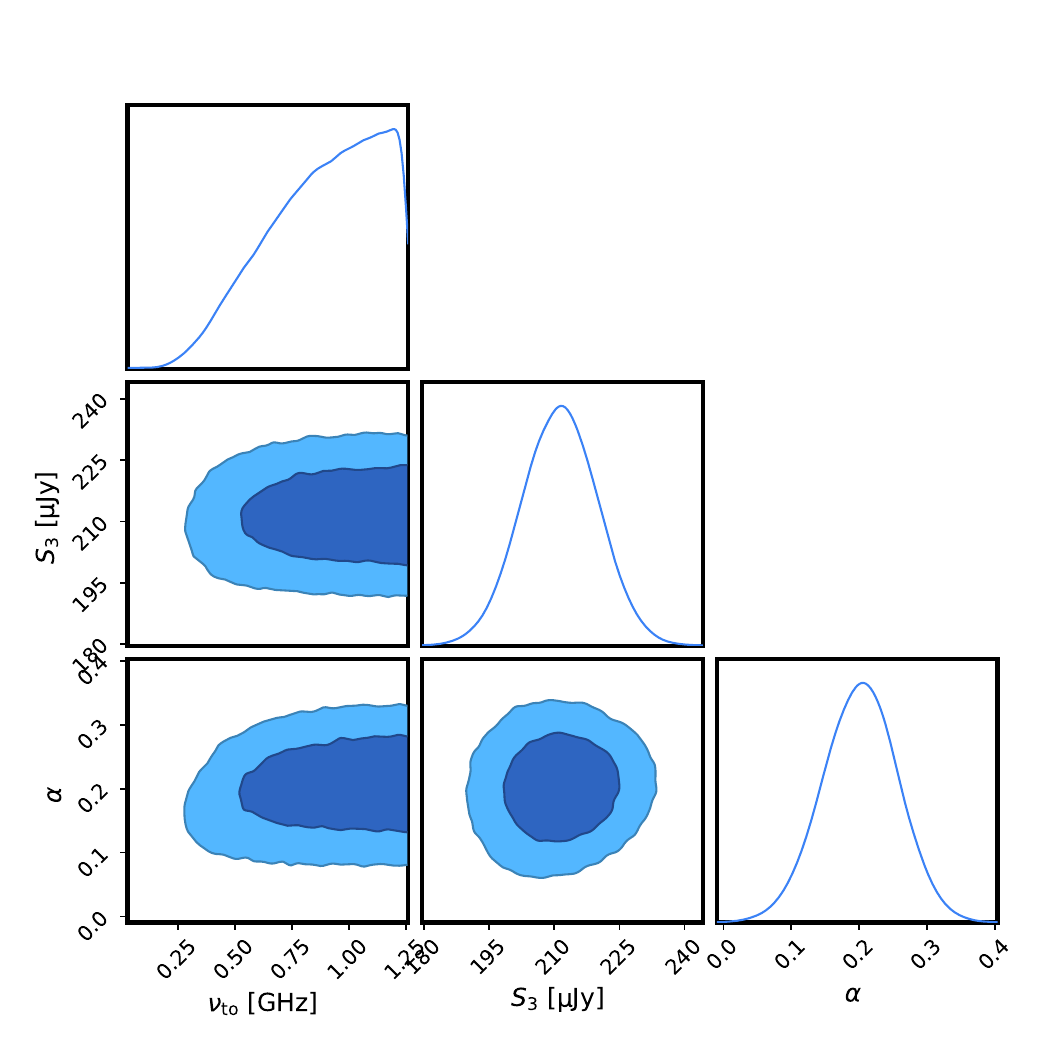}
    \caption{ Posterior probability distributions of the fitted parameters $(\nu_{\rm to},\; S_3, \; \alpha)$ for the spectrum shown in Fig. \ref{fig: S5_spectrum_break}. Dark and light areas refer to the $68\%$ and $95\%$ confidence levels, respectively.}
    \label{fig: corner}
\end{figure}

We fitted the model in Eq. \ref{eq: model spectrum} to all available flux density measurements and the 144 MHz upper limit following the MCMC method through the {\tt emcee}\footnote{\url{https://emcee.readthedocs.io/en/stable/}} package. The considered likelihood function includes the contributions from both measurements and upper limits, which is
\begin{equation}
\ln \mathcal L \equiv \ln \mathcal{L}(\mathcal{D}|\boldsymbol\vartheta)= \ln \mathcal{L}^{\rm meas} + \ln \mathcal{L}^{\rm ul},
\end{equation}
where $\boldsymbol \vartheta = {\{\nu_{\rm to}, S_3, \alpha}\}$ are the fitting parameters and $\mathcal{D} = (\nu_i, S_i)$ are the measured flux densities. We considered a standard Gaussian likelihood for detections, which is
\begin{equation}\label{eq: likelihood}
\ln \mathcal{L}^{\rm meas} =
- \frac{1}{2} \sum_{i} \left[
\frac{(S_i - S_{\rm mod}(\nu_i;\boldsymbol{\vartheta}))^2}{\sigma_i^2} + \ln(2\pi\sigma_i^2)
\right],
\end{equation}
where $S_{\rm mod}$ is the flux density predicted by the model and  $\sigma_i$ is the uncertainty on the measured flux density. The 144 MHz upper limit is taken into account through a likelihood term defined as the cumulative distribution function of a Gaussian (e.g. \citealt{Sawicki12,Yang20_ul,Stathopoulos24}), being:
\begin{equation}
    \ln \mathcal{L}^{\rm ul} = \ln \left\{
\frac{1}{2} \left[ 1 + \mathrm{erf} \left( \frac{S_{\rm ul} - S_{\rm mod}(\nu_{\rm ul};\boldsymbol{\vartheta})}{\sqrt{2}\,\sigma_{\rm ul}} \right) \right]
\right\} \;,
\end{equation}
where $\nu_{\rm ul}$ and $\sigma_{\rm ul}$ are the reference frequency and rms noise of LoTSS image (Fig. \ref{fig: surveys}). We considered flat
priors on $\boldsymbol{\vartheta}$, leaving the parameters free to vary in the ranges [10, 1260] MHz, [0, 1] Jy, [-2, 2] for $\nu_{\rm to}$, $S_3$, and $\alpha$, respectively. The resulting contour plots from the MCMC fit are shown in Fig. \ref{fig: corner}, while the fitted spectrum is shown in Fig. \ref{fig: S5_spectrum_break}.

We used the fitted parameters to constrain the size and age of the nebula in the MWN framework, following the expressions reported in \cite{Rahaman25}. We define $\hat{\nu}_{\rm sa} =\nu_{\rm sa}/{\rm 0.5 \; GHz}$, $\hat{\nu}_{\rm m} =\nu_{\rm m}/{\rm 2 \; GHz}$, $\hat{\nu}_{\rm c} =\nu_{\rm c}/{\rm 100 \; GHz}$, $\hat{L}_{\rm \nu_{\rm m}} =L_{\rm \nu_{\rm m}}/{\rm 2\cdot 10^{29} \; erg \; s^{-1} \; Hz^{-1}}$, $\hat{E}_{\rm SNR} =E_{\rm SNR}/{\rm 2\cdot 10^{50} \; erg}$, and $\hat{M}_{\rm *} =M_{*}/{10\; M_{\odot}}$, where $E_{\rm SNR}$ is the energy of the supernova remnant (SNR) embedding the nebula and $M_{\rm *}$ is the mass of the SNR progenitor. The radius and age are estimated as:
\begin{equation}
    R  = 0.06 \cdot  \hat{\nu}_{\rm sa}^{-\frac{5}{7}} \hat{\nu}_{\rm m}^{-\frac{5}{14}} \hat{\nu}_{\rm c}^{-\frac{13}{42}} \hat{L}_{\rm \nu_{\rm m}}^{\frac{3}{7}} \hat{E}_{\rm SNR}^{\frac{1}{14}}\hat{M}_{*}^{-\frac{1} {14}} \; \;  {\rm pc} \; \; ,
    \label{eq: radius}
\end{equation}
\begin{equation}
    t  = 37 \cdot  \hat{\nu}_{\rm sa}^{-\frac{5}{7}} \hat{\nu}_{\rm m}^{-\frac{5}{14}} \hat{\nu}_{\rm c}^{-\frac{13}{42}} \hat{L}_{\rm \nu_{\rm m}}^{\frac{3}{7}} \hat{E}_{\rm SNR}^{-\frac{3}{7}}\hat{M}_{*}^{\frac{3} {7}} \; \;  {\rm yr} \;\; .
    \label{eq: radius}
\end{equation}
As fiducial assumptions, we fixed the (unknown) $\nu_{\rm c}$, $E_{\rm SNR}$, and $M_{\rm *}$ parameters to their reference normalisation, and considered $\nu_{\rm sa}=200$ MHz. By assuming values of $\nu_{\rm m}= \nu_{\rm to}=370$ MHz and $\nu_{\rm c}=6$ GHz (i.e. their inferred lower limits), we obtain upper limits on the radius and age of the nebula of $R\leq 0.4$ pc and $t\leq 250$ yr. We notice that for a lower $\nu_{\rm sa}=100$ MHz, the upper limit on the age would not dramatically increase ($t\leq 400$ yr). For any combination of higher values of $\nu_{\rm sa}$, $\nu_{\rm m}$, and $\nu_{\rm c}$, the age would be lower.

\end{appendix}
\FloatBarrier

\end{document}